\documentclass[review]{elsarticle}

\journal{Physics Letters B}
\usepackage{xcolor, soul}
	\definecolor{darkgreen}{rgb}{0.0,0.55,0.0}
\usepackage{amssymb}
\usepackage[colorlinks=true, allcolors=darkgreen]{hyperref}
\usepackage{lineno}
	\modulolinenumbers[5]
\usepackage{empheq}
\usepackage{float}
\usepackage{cancel}

\def\beq#1\eeq{\begin{align}\begin{aligned} #1 \end{aligned}\end{align}}
\def\bseq#1\eseq{\begin{subequations}\begin{align} #1 \end{align}\end{subequations}}
\def\bseg#1\eseg{\begin{subequations}\begin{gather} #1 \end{gather}\end{subequations}}
\def\bbeq#1\ebeq{\begin{empheq}[box=\fbox]{align}\begin{aligned} #1 \end{aligned}\end{empheq}}
\def\be#1\ee{\begin{align} #1 \end{align}}
\def\bbeq#1\ebeq{\begin{empheq}[box=\fbox]{align} #1 \end{empheq}}
\def\beg#1\eeg{\begin{gather} #1 \end{gather}}

\renewcommand{\vec}{\mathbf}

\newcommand{\nc}{\newcommand}
\newcommand{\ba}{\begin{eqnarray}}
\newcommand{\ea}{\end{eqnarray}}
\newcommand{\ban}{\begin{eqnarray*}}
\newcommand{\ean}{\end{eqnarray*}}

\nc{\noind}{\noindent}

\newcommand{\sgn}{\text{sgn}}
\newcommand{\sqrts}{\sqrt{s}}
\newcommand{\lag}{\mathcal{L}}
\newcommand{\M}{\mathcal{M}}
\newcommand{\pd}{\partial}

\newcommand{\sigmav}{\langle\sigma v\rangle}

\newcommand{\wt}{\widetilde}

\newcommand{\gev}{\text{ GeV}}

\nc{\lsp}{\;\;\;\;}
\nc{\ad}{\lsp\text{and}\lsp}

\nc{\non}{\nonumber}
\def\lsim{\mathrel{\raise.3ex\hbox{$<$\kern-.75em\lower1ex\hbox{$\sim$}}}}
\def\gsim{\mathrel{\raise.3ex\hbox{$>$\kern-.75em\lower1ex\hbox{$\sim$}}}}

\begin{document}

\begin{frontmatter}

\title{$t$-channel singularities in cosmology and particle physics}

\author[fuw]{Bohdan Grzadkowski}
\author[fuw]{Micha{\l} Iglicki\corref{mail}}
\cortext[mail]{Corresponding author, email address: michal.iglicki@fuw.edu.pl}
\author[ujk,ncbj]{Stanis{\l}aw Mr{\'o}wczy{\'n}ski}

\address[fuw]{Faculty of Physics, University of Warsaw, ul. Pasteura 5, 02-093 Warsaw, Poland}
\address[ujk]{Institute of Physics, Jan Kochanowski University, ul. Uniwersytecka 7, 25-406 Kielce, Poland}
\address[ncbj]{National Centre for Nuclear Research, ul. Pasteura 7,  02-093  Warsaw, Poland}

\begin{abstract}
The $t$-channel singularity of a cross section for a binary $2 \to 2$ scattering process occurs when a particle exchanged in the $t$ channel is kinematically allowed to be on its mass-shell, that is when the process can be viewed as a sequence of two physical subprocesses: a two-body decay $1\to 2$ and an inverse decay $2 \to 1$. We derive conditions for the singularity to be present in a binary process. A class of divergent cross sections for Standard Model processes has been determined and illustrated by the weak analog of the Compton scattering $Ze^{-}\to Ze^{-}$.  After critically reviewing regularization methods proposed in literature, we discuss singular processes which occur in a medium composed of particles in thermal equilibrium. The medium is shown to regulate the singularities  naturally as particles acquire a finite width. We demonstrate a possible cosmological application by calculating thermally averaged cross section for an elastic scattering within a simple scalar model. The transition probability, which is divergent in vacuum, becomes finite when the process occurs in a thermal bath due to the imaginary self-energy of the $t$-channel mediator computed within the Keldysh-Schwinger formalism.
\end{abstract}


\date{December 23, 2021}

\end{frontmatter}


\section{Introduction}

Peierls noticed long ago \cite{Peierls:1961zz} that when the hadron binary process $\pi N^\star\to N^\star \pi$ is described using the Feynman diagram with a proton exchange in the $t$-channel, the kinematics does not exclude the exchange of the proton that is on its mass-shell. Then, the amplitude suffers from a divergence which also appears in a variety of binary reactions with unstable particles in the initial and final states. It was later realized \cite{Brayshaw:1978xt} that the divergence occurs not only in binary but in any process if there is a phase-space point such that an amplitude of the processes can be viewed as a product of on-mass-shell amplitudes of real processes. In case of the reaction $\mu^+\mu^- \to W^+ e^-  \bar{\nu}_e$, which has received some attention \cite{Ginzburg:1995bc, Melnikov:1996na, Melnikov:1996iu, Melnikov:1996ft, Dams:2002uy, Dams:2003gn} because of its importance for future $\mu^+\mu^-$ colliders, the two possible subprocesses are $\mu^- \to \nu_\mu e^- \bar{\nu}_e$ and $ \nu_\mu \mu^+ \to W^+$. The reaction $\mu^+\mu^- \to W^+ e^- \bar{\nu}_e$, which proceeds via the muon-neutrino exchange in the $t$-channel, is thus kinematically possible with on-mass-shell neutrino. Since the self-energy of neutrino or any other stable mediator is pure real at the mass-shell, the resummed propagator does not regulate the divergence. 

Even though a very existence of the singularities is an interesting phenomenon which is practically important in some specific cases, an awareness of the problem among high-energy physicists is rather limited. There are altogether only a few studies of the singularities, besides the publications mentioned above see \cite{Coleman:1965xm,Nowakowski:1993iu}.

Here we discuss the problem of $t$-channel divergences in detail. After reviewing two proposed regularization methods \cite{Ginzburg:1995bc, Melnikov:1996na, Melnikov:1996iu, Melnikov:1996ft, Dams:2002uy, Dams:2003gn} which, however, are either not satisfactory or not universally applicable, we study the divergences in the context we have encountered them, i.e., when singular amplitude occurs not in vacuum but in an environment of gas particles. Specifically, working on a multi-component dark matter (DM) \cite{Ahmed:2017dbb} we have found that for certain configurations of particle masses there exist singular transition probabilities which enter collision terms of Boltzmann equations. 

We show here that a presence of a medium naturally regularizes the $t$-channel singularities since mediating particles become quasi-particles, i.e., particles whose properties are modified by interaction with medium constituents. The modification includes a change of energy-momentum relation and an appearance of finite width -- the imaginary contribution to self-energy that remains finite on mass-shell. To demonstrate how the regularization works we perform simple one-loop calculations within the Keldysh-Schwinger formalism, see, e.g., \cite{Chou:1984es}, showing that the thermally averaged transition probability is finite as long as the system's temperature is non-zero. 

\section{t-channel singularity}

	\begin{figure}[h]\begin{center}
	\includegraphics[width=0.35\textwidth]{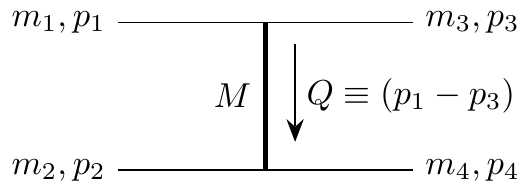}
       \vspace{-3mm}
	\caption{The considered $t$-channel mediated process}
	\label{fig:process}
	\end{center}\end{figure}

At the beginning let us discuss in some detail when the singularity occurs in a binary process with the $t$-channel mediator of mass $M$. The process is depicted in Fig.~\ref{fig:process}. We are looking for the condition in terms of masses and $s\equiv (p_1+p_2)^2$ that has to be satisfied for the singularity to appear. One easily finds that $t=M^2$ corresponds to the fixed angle between $\vec{p}_1$ and $\vec{p}_3$ given as
	\beq
	\cos(\theta_{13})=\frac{2E_1E_3-m_1^2-m_3^2+M^2}{2\sqrt{E_1^2-m_1^2}\sqrt{E_3^2-m_3^2}}\;.
	\eeq
The requirement $|\cos(\theta_{13})|\leq 1$ implies the following necessary and sufficient condition
	\beq
	\Delta\equiv\beta^2-4\alpha\gamma\ge 0 \ad
	\sqrt{s_1}\le\sqrts\le\sqrt{s_2} ,
	\label{sing_cond}
	\eeq
where $s_{1,2} \equiv (-\beta \mp \sqrt\Delta)/(2\alpha)$, and 
	\beg
	\alpha\equiv M^2\;,\qquad
	\beta\equiv M^4-M^2(m_1^2+m_2^2+m_3^2+m_4^2)
		+(m_1^2-m_3^2)(m_2^2-m_4^2)\;,
	\notag\\
	\gamma\equiv M^2(m_1^2-m_2^2)(m_3^2-m_4^2)
		+(m_1^2m_4^2-m_2^2m_3^2)(m_1^2-m_2^2-m_3^2+m_4^2)\;.
	\notag
	\eeg
It is also worth to specialize to the case of elastic scattering. When $m_1=m_3$ and $m_2=m_4$ we have $E_1=E_3$ and $|\vec p_1|=|\vec p_3|$ in the center-of-mass frame, so $t\equiv(\vec p_1-\vec p_3)^2=2m_1^2-2E_1^2+2|\vec p_1|^2\cos(\theta_{13})\le 0$. Hence, no massive mediator ($M > 0$) can be on-mass-shell. However, in the exchange channel, when $m_1=m_4$ and $m_2=m_3$, the singularity is possible. In a special case of the process $\nu_e W^+ \to W^+ \nu_e$ the condition (\ref{sing_cond}) was derived in \cite{Nowakowski:1993iu}.

The condition (\ref{sing_cond}) can be reformulated by saying that exactly one particle in the initial and in the final state must be unstable and the other must be stable. Then, there exists a range of $s$ that satisfies the equality (\ref{sing_cond}). Here, the terms `unstable' and `stable' are restricted to the interactions shown in Fig.~\ref{fig:process} only. If in addition the mediator called $\Phi$ is stable (in the standard sense), so that its width vanishes then the resummation of self-energies does not regularize the singularity.

\section{Relevance of the singularity}

There is a variety of Standard Model processes plagued by the divergence. We mention here the elastic scattering  $W^- e^- \to e^- W^-$  mediated by the $\nu_e$ exchange as well as $Z e^- \to e^- Z$ and $W^+ \nu_e \to \nu_e W^+$ both with $e$ being the mediator. The examples of inelastic processes are $Z \nu_e \to e^- W^+$ and $W^- \nu_e \to e^- Z$ mediated by $e^+$ and $\bar{\nu}_e$, respectively. Those processes can take place in collider scattering or during evolution of the early Universe.

Let us look closer at the $Z e^- \to e^- Z$ which is a weak analog of the Compton scattering with $\gamma$ being replaced by $Z$ boson. In this case, $s_1$ and $s_2$, which enter the condition  (\ref{sing_cond}), are $s_1\equiv 2m_Z^2+m_e^2$ and $s_2\equiv \frac{(m_Z^2-m_e^2)^2}{m_e^2}$. In between $s_1$ and $s_2$ the cross section diverges. The length of the region is $s_2-s_1=m_Z^4/m_e^2(1-4m_e^2/m_Z^2)\gg m_Z^2$.

	\begin{figure}[H]\begin{center}
	\includegraphics[width=0.6\linewidth]{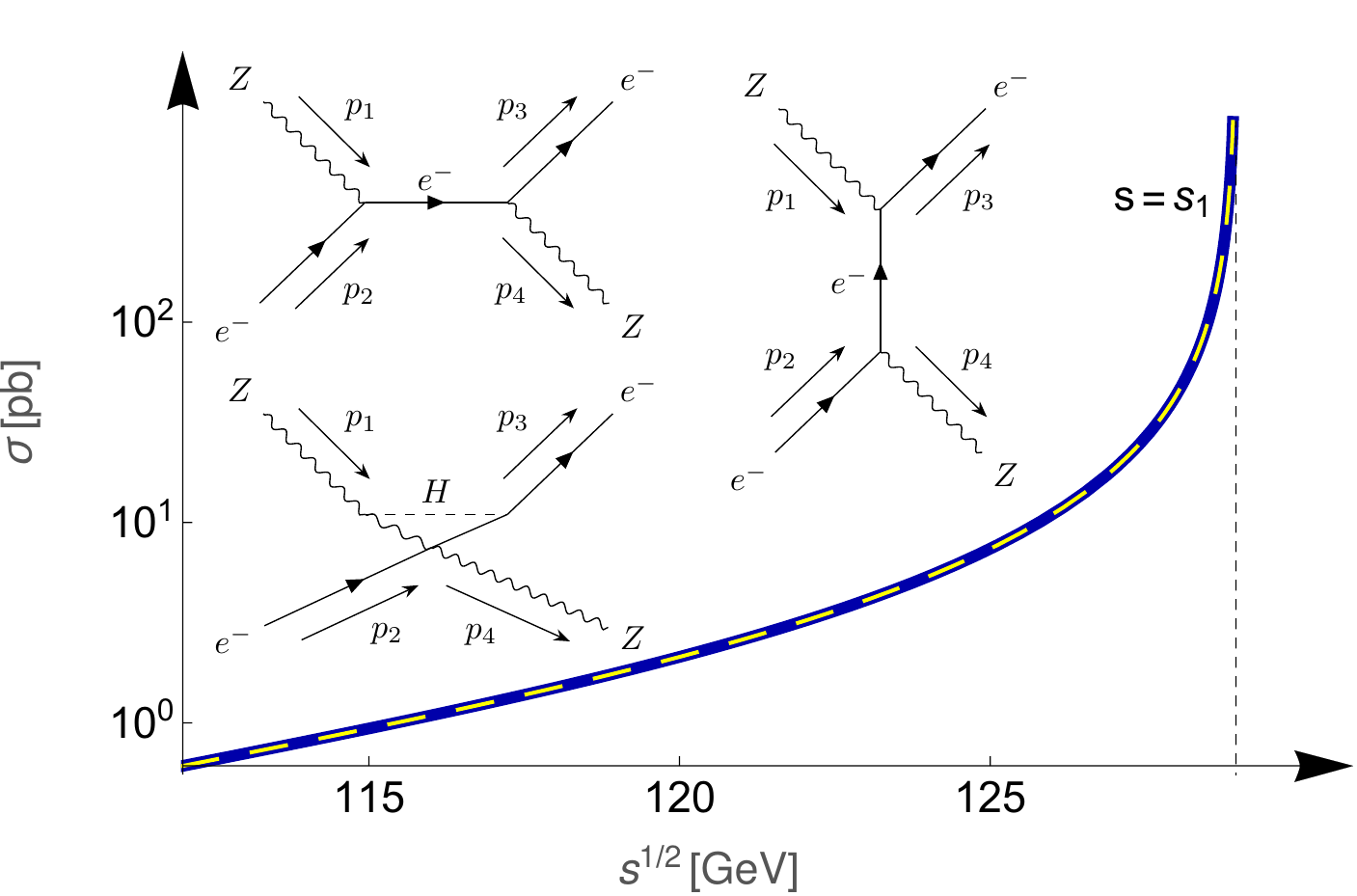}
	\caption{The solid blue line shows numerically integrated total cross section $\sigma_{Ze\to eZ}$ as a function of $\sqrt{s}$. The dashed yellow line represents the integrated Klein-Nishina-like formula for $m_e=0$. The contributing Feynman diagrams are also depicted. The divergent region starts at $\sqrt{s} \simeq \sqrt{2}m_Z \simeq 129$~GeV, it ends at $\sqrt{s}=\sqrt{s_2}\simeq m_Z^2/m_e\approx 1.6 \cdot 10^7$~GeV that is beyond the range of shown $\sqrt{s}$.}
	\label{fig:Ze-eZ-5GeV-sqrts}
	\end{center}\end{figure}

In Fig.~\ref{fig:Ze-eZ-5GeV-sqrts} we plot the numerically computed total cross section $\sigma_{Ze\to eZ}$ as a function of $\sqrt{s}$. The contributing Feynman diagrams are depicted in the figure. The divergent region of $\sqrt{s}$ that starts at $\sqrt{s}=\sqrt{s_1}\approx \sqrt{2}m_Z  \approx 129$~GeV is clearly seen. The right edge of the divergent region located at $\sqrt{s}=\sqrt{s_2}\simeq m_Z^2/m_e \approx 1.6 \cdot 10^7$~GeV is not shown.

Since the electron mass $m_e$ is much smaller than the $Z$ boson mass $m_Z$, the numerically found curve shown in Fig.~\ref{fig:Ze-eZ-5GeV-sqrts} can be reproduced by the cross section of $Z e^- \to e^- Z$ computed analytically for $m_e =0$.  The analog of the Klein-Nishina formula~\cite{Klein:1929ab} for the weak Compton scattering applicable for $m_Z^2 \lsim s \lsim 2m_Z^2$ reads:
	\begin{align}\label{KNform}\begin{aligned}
	\frac{d\sigma_{Ze\to eZ}}{dt} &= g^4\,\frac{g_v^4+6g_a^2g_v^2+g_a^4}{12\,\pi\,c_w^4} \frac{(m_Z^4-s\,t)}{(s-m_Z^2)^2}\\
	&\mkern20mu\times \frac{m_Z^4\,s^2+4m_Z^2(2m_Z^2-s)\,st+(m_Z^4-4m_Z^2s+s^2)\,t^2}{m_Z^4\,s^2\,t^2} ,
	\end{aligned}\end{align}
where $g$ is the $SU(2)$ gauge coupling constant, $c_w \equiv \cos\theta_W$, $g_v\equiv (1-4 \sin^2\theta_W)/4$ and $g_a\equiv 1/4$. We note that in the limit $m_e\to 0$ the $t$-channel divergence appears at $t=0$, $s_1=2m_Z^2$ and $s_2=\infty$. The total cross section shown by the yellow dashed line in Fig.~\ref{fig:Ze-eZ-5GeV-sqrts}, corresponds to
$\sigma_{Ze\to eZ}(s) = \int_{2m_Z^2-s}^{m_Z^4/s}\frac{d\sigma}{dt}\,dt$.

As we also already mentioned, the analysis of the process $\pi N^\star\to N^\star \pi$ led to the discovery of $t$-channel singularities \cite{Peierls:1961zz}. Numerous unstable hadron species provide many examples of singular hadronic processes whenever they could be modeled in terms of Feynman diagrams. The processes play an important role in relativistic heavy-ion collisions where numerous hadrons, which are produced, can be treated as a hadron gas \cite{Florkowski:2010zz}. Due to frequent hadron-hadron collisions the gas is in thermal equilibrium before the freeze-out. Since a proton is the only stable hadron, only processes with the proton exchange can suffer, strictly speaking, from the $t$-channel singularity. However, if a process with, say, $\pi^+$ exchange is regularized with its width controlled by weak interactions, the cross section becomes very big and the problem is still there.

\section{Regularization by the mediator width}

Let us now discuss how the issue of $t$-channel singularities has been addressed. The first natural possibility is to regularize the divergence by a mediator width using the Breit-Wigner propagator with $M^2$ replaced by $M^2-iM\Gamma$. Then, the $t$-channel amplitude is proportional to $(t-M^2+iM\Gamma)^{-1}$, it is finite at $t=M^2$ and the cross section is proportional to $1/\Gamma$. The cross section is obviously infinite if the mediator is truly stable and $\Gamma=0$. However, it is also abnormally big if a process of strong interactions is regularized by the width driven by weak interactions. Consequently, the cross section of a hadronic process with, e.g., $\pi^+$ exchange becomes orders of magnitude bigger than the typical hadronic cross section at $t=M^2$. So, it is a doubtful regularization technique. 

\section{Stable mediator -- known approaches}
Ginzburg \cite{Ginzburg:1995bc} proposed another method which originates from the observation that some initial-state particles are unstable. Assuming that the particle `1' decays as $1\to \Phi + 3$ with the width $\Gamma_1$, the particle wave function in the rest frame should include, according to the method \cite{Ginzburg:1995bc}, the factor 
	\beq
	e^{-im_1t}e^{-\Gamma_1 t}=e^{-i(m_1-i\Gamma_1)t}\;.
	\eeq
Hence, we can treat $\wt{m}_1\equiv m_1 - i\Gamma_1$ as a complex mass of the particle `1'. After a Lorentz boost, its four-momentum is also complex
	\beq
	(\wt E_1,\wt{\vec p}_1)=\Big(1-i\frac{\Gamma_1}{m_1}\Big) (E_1,\vec p_1)\;.
	\eeq
Consequently, the variable $t$ is complex as well
	\beq
	t &=(\wt{E}_1-E_3)^2-(\wt{\vec p}_1-\vec p_3)^2\\
	  &=\Big(E_1-E_3-i\frac{\Gamma_1}{m_1}E_1\Big)^2
		-\Big(\vec p_1-\vec p_3 - i \frac{\Gamma_1}{m_1}\vec p_1\Big)^2\;,
	\eeq
and the singularity condition $t=M^2$ is never satisfied. Unfortunately, the Ginsburg's method contradicts the exact energy-momentum conservation. Indeed, the real part of $t$ is 
	\beq\label{eq:GinzRet13}
	\Re t
	&=(E_1-E_3)^2-(\vec p_1-\vec p_3)^2-\Gamma_1^2\;,
	\eeq
but using $p_1-p_3=p_4-p_2$ and observing that the particle `4' is also unstable with the width $\Gamma_4$, one gets the expression (\ref{eq:GinzRet13}) with $\Gamma_1$ replaced by $\Gamma_4$.  Since, in general, $\Gamma_1 \not=\Gamma_4$, there appears an energy-momentum mismatch. Therefore, the complex mass solution is rather unsatisfactory. 


Alternative method suggested by Melnikov and Serbo \cite{Melnikov:1996na} and further developed in \cite{Melnikov:1996iu, Melnikov:1996ft}, see also \cite{Dams:2002uy, Dams:2003gn}, makes use of a~finite size of a beam of long-living particles. The method was worked out in the context of a future $\mu^+\mu^-$ collider. The cross section is then proportional not to the modulus squared of matrix element $|\M_{fi}|^2$ but to $\M_{fi}^*\M_{fi'}$ where the final state is the same but initial momenta slightly differ. The difference is of the order of $1/a$ with $a$ being the inverse transverse size of the initial beam. Consequently, the singular points of the amplitude $\M_{fi}$ and $\M_{fi'}$ are slightly shifted from each other and the divergent integral is changed into 
	\beq
	\int\frac{dt}{|t-M^2 + i \varepsilon|^2}
	\quad\rightarrow\quad
	\int\frac{dt}{(t - \Pi - M^2+ i \varepsilon)(t + \Pi -M^2 - i \varepsilon)} \;,
	\eeq
where the integration is performed over a vicinity of the point $t=M^2$ and $\Pi = m/a$ with $m$ being the mass of incoming particles. For a symmetric Gaussian beam profile $\sqrt{2/\pi}\,a$ is equal to the Gaussian width of the beam.  One shows that the regularized integral is proportional to $1/\Pi = a/m$. 

The beam-size method has a solid foundation and is applicable to a collider experimental situation. However, it is not suitable if a singular process occurs in a~particle gas as in the cosmological context. In this case transition probabilities that enter Boltzmann equations describe scatterings of statistically distributed particles moving at various directions and energy-momenta.

\section{Thermal-field-theory regularization}

It turns out that particles propagating and interacting with the gas become quasi-particles because of their finite mean free path or mean free flight time. Particle self-energies, which are pure real at mass-shell in vacuum, acquire imaginary contributions in medium. Consequently, quasi-particles have finite widths which allow to regularize singular amplitudes. We emphasize that in a medium there is no fundamental difference between particles that are unstable in vacuum and those that are stable, as all particles are actually quasi-particles of finite width.

We are going to demonstrate that mechanism using the Keldysh-Schwinger formalism, see, e.g., \cite{Chou:1984es}, applicable to both equilibrium and non-equilibrium systems of quantum fields. A toy model that we adopt for illustrative calculations consists of three real scalar fields  $\varphi_{1,2}$ and $\Phi$. The Lagrangian reads
	\beq
	\lag
	&=\phantom{+}\frac12\left[(\pd^\mu\varphi_1)(\pd_\mu\varphi_1)-m_1^2\varphi_1^2\right]
		+\frac12\left[(\pd^\mu\varphi_2)(\pd_\mu\varphi_2)-m_2^2\varphi_2^2\right]\\
	&\phantom{=}+\frac12\left[(\pd^\mu\Phi)(\pd_\mu\Phi)-M^2\Phi^2\right]
		\;\;+\mu\,\varphi_1\varphi_2\Phi\;,
	\eeq
where masses of $\varphi_1$, $\varphi_2$ and $\Phi$ are denoted as $m_1$, $m_2$ and $M$, respectively, and $\mu>0$ is an interaction coupling of mass dimension. We assume that $m_1 > m_2 + M$, so that $\varphi_2$ and $\Phi$ are stable (in cosmological applications they would be candidates for DM) while $\varphi_1$ decays as $\varphi_1\to \varphi_2 \Phi$. Consequently, the amplitude of the process  $\varphi_1\, \varphi_2 \to \varphi_2 \, \varphi_1$ mediated by $\Phi$ in $t$ channel is singular.

	\begin{figure}[h]
	\begin{center}
	\includegraphics[width=0.3\linewidth]{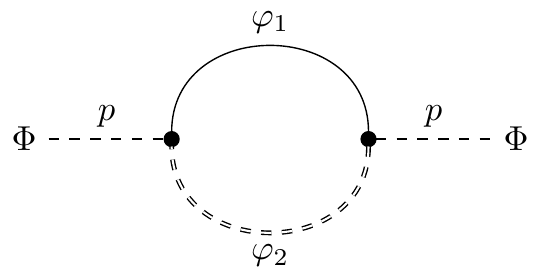}
	\vspace{-3mm}
	\caption{The one-loop diagram for the self energy of $\Phi$}
	\label{fig:1loop}
	\end{center}
	\end{figure}

We are going to compute the one-loop self-energy of the field $\Phi$ represented by the graph shown in Fig.~\ref{fig:1loop}. Although our reasoning does not require the assumption of thermal equilibrium and the Keldysh-Schwinger formalism works for equilibrium and non-equilibrium systems, we consider the quantum fields $\varphi_1$, $\varphi_2$ and $\Phi$ in global thermodynamic equilibrium to simplify the problem. So, a quasi-particle $\Phi$ propagates across the thermal bath of temperature $T$ interacting with the system's constituents. 

We are interested in a nonzero imaginary contribution to $\Phi$ self-energy at the mass-shell, as it is needed to regularize the singular transition probability. The contribution usually corresponds to a physical scattering process and it appears at the two-loop level when binary interactions contribute to the self-energy. However, here the contribution of interest appears already at the one-loop level due to the decay $\varphi_1\to \varphi_2 \Phi$. 

For the $t$-channel singularity to appear, particles unstable in vacuum must be present both in the initial and in the final states. Therefore an intuition suggests that a solution of the problem should be related to the very same instability, that, on the other hand, implies a lack of proper asymptotic in and out states in vacuum. It will be shown below that indeed the treatment proposed here is intrinsically related to the instability. It is worth mentioning that in general, in statistical quantum field theory, there are no asymptotic states regardless if they are made of states that are stable or unstable in vacuum since in a remote past and in a remote future all particles do interact with their environment.

As explained in detail in, e.g., \cite{Czajka:2010zh}, the one-loop retarded self-energy $\Pi^+(p,T)$ we are interested in is
	\beq
	\label{Pi-p-1}
	\Pi^+(p,T) = \frac{i\mu^2}{2}
	\!\!\int\!\! \frac{d^4k}{(2\pi)^4} 
       \Big[\Delta_1^+(k+p) \, \Delta_2^{\rm sym} (k,T)
       + \Delta_1^{\rm sym}(k,T) \, \Delta_2^-(k-p) \Big],
	\eeq
where the indices 1 and 2 refer to the fields  $\varphi_1$ and $\varphi_2$. The free Green's functions $\Delta^\pm (p)$ and $\Delta^{\rm sym}(p,T)$ of a~scalar field of mass $m$ are
	\bseq 
       \label{D-pm}
	&\Delta^{\pm}(p) = \frac{1}{p^2 - m^2 \pm i\, {\rm sgn}(p_0)\,\varepsilon},
      \\[2mm]
	&\begin{aligned} \label{D-sym}
	\Delta^{\rm sym}(p,T) &= -\frac{i\pi}{E_p}\Big( \delta (E_p - p_0) + \delta (E_p + p_0)\Big)
	\big[2 f(E_p,T)+ 1\big]\;,
	\end{aligned}
	\eseq
where $E_p \equiv \sqrt{\vec p^2 + m^2}$ and $f(E_p,T)$ is the Bose-Einstein distribution function $f(E_p,T) = (e^{\beta E_p} -1)^{-1}$ with $\beta \equiv 1/T$. 

Our calculations are performed in the rest frame of the heat bath which in cosmology corresponds to the reference frame where the Universe is homogeneous and isotropic. However, after replacing $\beta E_p$ by $\beta u^\mu p_\mu$ our calculations would be valid in a reference frame which moves with respect to the heat bath with the four-velocity $u^\mu$.  

Substituting the functions (\ref{D-pm}) and (\ref{D-sym}) into Eq.~(\ref{Pi-p-1}), the imaginary part $\Sigma(\vec p,T) \equiv \Im\Pi^+(p,T)$ for the on-mass-shell momentum, i.e., for $p_0 = E_p \equiv \sqrt{\vec p^2+M^2}$, is obtained as
	\beq
	\label{ImPi-2}
	\Sigma(\vec p,T) &= - \frac{\pi}{4} \mu^2 
		 \int \frac{d^3k}{(2\pi)^3}\bigg\{ 
		\frac{e^{\beta E_{k2}} +1 }{e^{\beta E_{k2}} -1 }\,\\
	&\mkern10mu\times
		\bigg[
		\frac{\sgn\big(E_{k2} + E_p\big)}{E_{k2}}\,
		\delta \Big( M^2+m_2^2 - m_1^2 + 2E_{k2}E_p - 2\vec k \cdot \vec p\Big)
		\\[2mm]
	&\mkern30mu+
		\frac{\sgn\big(-E_{k2} + E_p\big)}{E_{k2}}\,
		\delta \Big( M^2+m_2^2 - m_1^2 - 2 E_{k2} E_p - 2 \vec k \cdot \vec p\Big)
		\bigg]\\
	&\mkern10mu+
		[1\leftrightarrow 2\,,\;\vec k\to -\vec k]
	\bigg\}\;,
	\eeq
where $E_{k1}\equiv\sqrt{\vec k^2 + m_1^2}$ and $E_{k2}\equiv\sqrt{\vec k^2 + m_2^2}$. After tedious calculations the following result has been found
	\beq\label{ImPi-7}
	&\Sigma(|\vec p|,T)
	= - \frac{\mu^2 }{16\pi}\;\frac{1}{\beta|\vec p|}
	\times
	\Bigg[
	\ln\frac{e^{\beta(A+C)}-1}{e^{\beta A}-1}
	-\ln\frac{e^{\beta(A+B+C)}-1}{e^{\beta(A+B)}-1}
	\Bigg]\;,
	\eeq
where
	\bseg
	A\equiv\frac{(m_1^2-m_2^2-M^2)E_p - 2Mk_*\,|\vec p|}{2M^2}, \notag\\
	B\equiv E_p, \quad C\equiv\frac{2\,k_*\,|\vec p|}{M}, \quad  k_* \equiv\frac{\sqrt{\lambda(m_1^2,m_2^2,M^2)}}{2M}\;,\notag
	\eseg
and $\lambda(a,b,c)\equiv a^2+b^2+c^2-2ab-2ac-2bc$. Since the system is isotropic the self-energy depends only on $|\vec p|$ not $\vec p$. 
Note that the imaginary part of the self-energy (\ref{ImPi-7}) vanishes in the zero-temperature limit.

\begin{figure}[h]\begin{center}
	\includegraphics[width=0.8\linewidth]{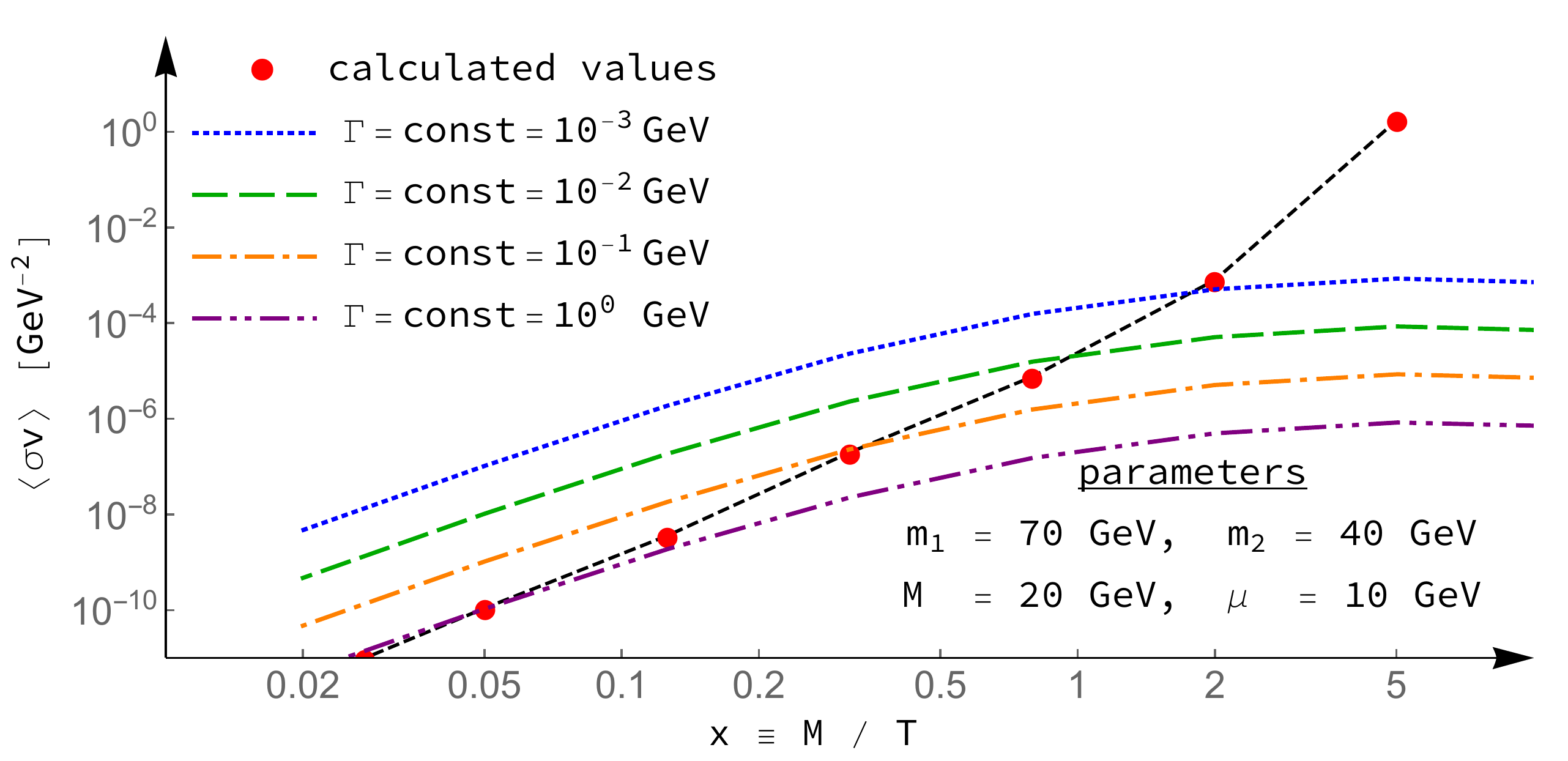}
	\caption{The thermally averaged cross section for the process $\varphi_1 \, \varphi_2 \to \varphi_2 \, \varphi_1$. The red points show results of Monte Carlo phase-space integration in (\ref{thermal-cross}) and the black dashed line is a polynomial fit drawn to guide reader's eyes. The blue short-dashed, green dashed, yellow dashed-dotted and purple dashed-double-dotted lines show results obtained adopting the Breit-Wigner formula with a constant width $\Gamma(|\vec p_{13}|,T)=10^{-4}, 10^{-3}, 10^{-2}, 10^{-1} \gev$, respectively.}
	\label{fig:sigmav}
	\end{center}\end{figure}

In Fig.~\ref{fig:sigmav} we present the thermally averaged cross section for $\varphi_1\, \varphi_2 \to \varphi_2 \, \varphi_1$ process mediated by $\Phi$ in the $t$ channel. This quantity is defined as
	\beq
       \label{thermal-cross}
	&\sigmav(T)
      =\mu^4\int d\Pi_1\,d\Pi_2\,\frac{f_1(E_1,T)}{n_1(T)}\,\frac{f_2(E_2,T)}{n_2(T)}\\
	&\mkern80mu \times
	\int d\Pi_3\,d\Pi_4\,\frac{(2\pi)^4\,\delta^{(4)}(p_1+p_2-p_3-p_4)}{(t-M^2)^2+\Sigma^2(|\vec p_{13}|,T)}
	\;,
	\eeq
where $\vec p_{13}\equiv \vec p_1-\vec p_3$, while $f_k(E_k,T)$ and $n_k(T)$ are, respectively, the Bose-Einstein distribution function and the equilibrium number density of particle species $\varphi_k$ ($k=1,2$), and ${d\Pi_i\equiv d^3p_k/[(2\pi)^3\,2E_k]}$ denotes a~phase-space element for $i=1,2,3,4$. The parameters are chosen as $m_1 = 70$ GeV,  $m_2 = 40\gev$, $M = 20\gev$, and $\mu = 10\gev$ which are of the order of magnitude relevant for the DM studies \cite{Ahmed:2017dbb}. Strictly speaking, the thermally averaged cross section is not a cross section as it does not describe transitions between asymptotic states. However, it approximates the cross section in the limit of a dilute system.

Eq.~(\ref{thermal-cross}) clearly shows that the thermally averaged cross section is divergent if the equality $t=M^2$ is kinematically allowed and $\Sigma^2(|\vec p_{13}|,T)$ is absent. The integration variables can be changed to explicitly integrate over $s$, with the lower integration limit $s_{\text{min}}\equiv\text{max}[(m_1+m_2)^2,(m_3+m_4)^2]$. Therefore, whenever $s_{\text{min}}<s_2$ and $\Delta > 0$, the integration hits the divergent region corresponding to $t=M^2$, so that the integrand diverges in some range of $s$. The non-zero imaginary part of self-energy removes the divergence. We note that the one-loop self energy includes not only the imaginary but also the real part which shifts the mass $M$. However, the shift is irrelevant for our considerations so the real part of self-energy is neglected. 

As seen in Fig.~\ref{fig:sigmav} the thermally averaged cross section quickly increases with ${x\equiv M/T}$. It is a consequence of the thermal width $\Gamma(|\vec p_{13}|,T) \equiv \Sigma(|\vec p_{13}|,T)/M$ diminishing with decreasing temperature. In the limit of zero temperature the thermal width vanishes and the cross section blows up. For comparison we have also shown the cross section calculated for a few selected constant widths. It is worth noticing that for small $x$, i.e., $x \lsim 0.1$,  $\Gamma(|\vec p_{13}|,T)$ could be approximated by a constant width $\Gamma(|\vec p_{13}|,T) \simeq 0.1\gev$ using the Breit-Wigner formula.

\section{Summary}
The $t$-channel singularities have been discussed paying particular attention to binary processes. We have formulated conditions for the singularity to appear and the singular Standard Model process $Ze^- \to e^- Z$ has been 
discussed in detail. After reviewing critically regularization methods proposed in literature we have focused on singular processes which occur in a medium of gas of particles.  It has been argued that the medium naturally regulates the singularity as particles become quasi-particles with a finite width. As an illustration we have computed, using the Keldysh-Schwinger formalism, the imaginary part of the self-energy of a scalar mediator and a thermally averaged cross section for the process, which is singular in vacuum. It has been shown that the cross section becomes finite when the process occurs in the thermal bath. 
It should be emphasized that even if the singularity is regularized by the thermal width, the cross section is still large. The singularities appear naturally in multicomponent dark matter scenarios, e.g., in the vector-fermion model discussed in \cite{Ahmed:2017dbb}, so they could be adopted in models of strongly interacting dark matter as well, see \cite{Duch:2019vjg,Duch:2018ucs,Duch:2017nbe} for corresponding zero-temperature mechanism of enhancement emerging from $s$-channel mediator exchange.

\vspace{3mm}

The authors thank Jose Wudka for his interest at the starting stage of this project and Jacek Pawe{\l}czyk for discussions. 
This work was partially supported  by the National Science Centre (Poland) under grants 2017/25/B/ST2/00191 and 2020/37/B/ST2/02746 in case of B.G. and M.I., and under grant 2018/29/B/ST2/00646 in case of St.M.

\bibliography{citations}

\end{document}